\documentclass[a4paper,11pt]{article}
\pdfoutput=1 

\usepackage{jinstpub} 
\usepackage{lineno,hyperref}
\usepackage{epstopdf}
\usepackage{graphicx}
\usepackage{csquotes}

\title{\boldmath MPGD-based photon detectors for the upgrade of COMPASS RICH-1 and beyond}

\author{
	J.~Agarwala$^{1,b}$,
	M.~Alexeev$^4$,
	C.D.R.~Azevedo$^8$,
	F.~Bradamante$^{11}$,
	A.~Bressan$^{11}$,
	M.~B\"uchele$^7$,
	C.~Chatterjee$^{11}$,
	M.~Chiosso$^{10}$,
	A.~Cicuttin$^1$,
	P.~Ciliberti$^5$,
	M.L.~Crespo$^1$
	S.~Dalla~Torre$^5$,
	S.~Dasgupta$^{5,a}$
	O.~Denisov$^4$,
	M.~Finger$^2$,
	M.~Finger~Jr$^2$,
	H.~Fischer$^7$,
	L.~García~Ordóñez$^{1,3,11}$,
	M.~Gregori$^5$,
	G.~Hamar$^{5,c}$,
	F.~Herrmann$^7$,
	S.~Levorato$^5$,
	A.~Martin$^{11}$,
	G.~Menon$^5$,
	D.~Panzieri$^9$,
	G.~Sbrizzai$^{11}$,
	S. Schopferer$^7$,
	M. Slunecka$^2$,
	M. Sulc$^6$,
	F.~Tessarotto$^5$,
	J.F.C.A.~Veloso$^8$,
	Y.X.~Zhao$^{5,d}$
}

\affiliation{	$^1		$Abdus Salam ICTP and INFN Trieste, Trieste, Italy} 
\affiliation{	$^2		$Charles University, Prague, Czech Republic and JINR, Dubna, Russia} 
\affiliation{	$^3	    $Engineering Dep. of Trieste University, Trieste, Italy.} 
\affiliation{	$^4		$INFN Torino, Torino, Italy} 
\affiliation{	$^5		$INFN Trieste, Trieste, Italy} 
\affiliation{	$^6	    $Technical University of Liberec, Liberec, Czech Republic} 
\affiliation{	$^7		$Universit\"at Freiburg, Freiburg, Germany} 
\affiliation{	$^8		$I3N, Department of Physiccs, University of Aveiro, Aveiro, Portugal} 
\affiliation{	$^9		$University of East Piemonte, Alessandria and INFN Torino, Torino, Italy} 
\affiliation{	$^{10}		$University of Torino and INFN Torino, Torino, Italy} 
\affiliation{	$^{11}		$University of Trieste and INFN Trieste, Trieste, Italy} 

\affiliation{$^a$Corresponding author}
\affiliation{$^b$Present address: University of Pavia and INFN Pavia, Pavia, Italy}
\affiliation{$^c$Present address: Wigner Research Centre for Physics, Budapest, Hungary}
\affiliation{$^d$Present address: Institute of Modern Physics, Chinese Academy of Sciences, Lanzhou, 730000, China}

\emailAdd{shuddha.dasgupta@ts.infn.it}

\abstract{COMPASS is a fixed target experiment at CERN SPS aimed to study hadron structure and spectroscopy. Hadron identification in the momentum range between $3$ and $55 GeV/c$ is provided by a large gaseous Ring Imaging Cherenkov Counter, RICH-1. To cope with the challenges imposed by the new physics program of COMPASS, RICH-1 has been upgraded by replacing four MWPC-based photon detectors with newly developed MPGD-based photon detectors. The architecture of the novel detectors is a hybrid combination of two layers of THGEMs and a Micromegas. The top of the first THGEM is coated with CsI acting as a reflective photo-cathode. The anode is segmented in pads capacitively coupled to the APV-25 based readout. The new hybrid detectors have been commissioned during the 2016 COMPASS data taking and stably operated during the 2017 run. 
	\par
	In this paper design, construction, operation and performance aspects of the novel photon detectors for COMPASS RICH-1 are discussed.}


\arxivnumber{1234.56789} 

\proceeding{International conference on Instrumentation for Colliding Beam Physics  INSTR$\rq20$\\
	24$^{Th}$ February, 2020 to 28$^{Th}$ February, 2020\\Budkar Institute of Nuclear Physics, Novosibirsk, Russia}

\begin{document}
	\maketitle
	\flushbottom
	
	
	
	\section{Introduction}
	
	\par
\begin{figure}[h]
		\centering
		\includegraphics[width=0.8\linewidth]{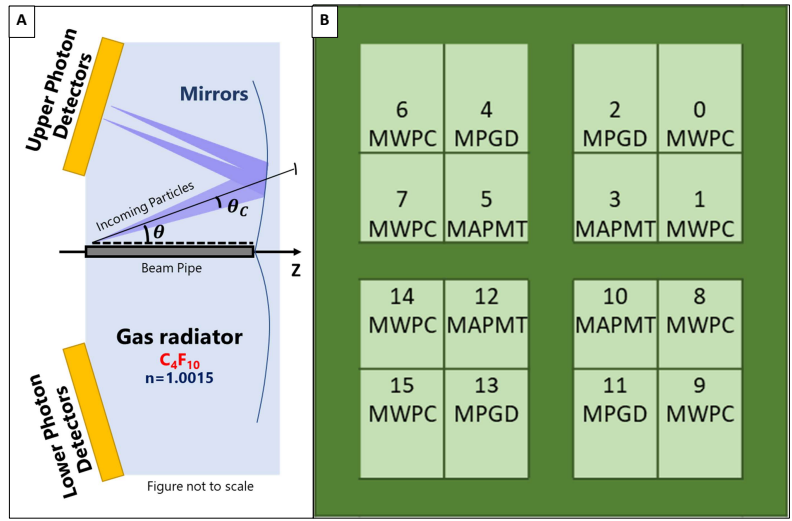}
		\caption{A. The Cherenkov photon propagation and focusing: the principle is illustrated by a schematic side view. B. Photon detector arrangement (not to scale), where the different detection technologies in use are indicated.}
		\label{fig:richPrinciple}
	\end{figure}

 
	RICH-1~\cite{Albrecht} is a large gaseous Ring Image Cherenkov (RICH) Counter  providing Particle Identification (PID) for hadrons within the momentum range from $3$ to $55~ GeV/c$ for the COMPASS~\cite{compass} Experiment at CERN SPS. It consists of a 3 m long $C_{4}F_{10}$ gaseous radiator, where Cherenkov photons are generated when ultra-relativistic charged particles cross it; VUV spherical mirrors covering a surface of 21 $m^2$ reflect the photons and focus them onto a 5.5 $m^2$ photo-detection surface sensitive to single photons (Fig\ref{fig:richPrinciple}-A). 
	Three photo-detection technologies are used in RICH-1: Multi Wire Proportional Chambers (MWPCs) with CsI photo-cathodes, Multi Anode Photo-Multipliers Tubes (MAPMTs) and novel Micro Pattern Gaseous Detector (MPGD) based Photon Detectors (PDs) (Fig.\ref{fig:richPrinciple}-B). The novel PDs are the focus of this article.
	\par
	RICH-1 was designed and built between 1996-2000, commissioned in 2001-2002 and is in operation since 2002. The whole photo-detection surface was originally equipped  with 16 MWPCs with CsI photo-cathodes of $\sim 600\times600$ mm$^2$ active area. In-spite of their good performance, MWPCs have limitations in terms of time resolution, maximum effective gain ($\sim 10^4$), time response($\sim \mu$s), rate capability and ageing of the CsI photo-cathodes. In 2006, four central chambers were replaced with detectors consisting of MAPMTs coupled to individual fused silica lens telescopes to cope, thanks to their excellent time resolution,  with the high particle rates of the central region. In parallel, an extensive R\&D program~\cite{THGEM_rd} aimed to develop MPGD-based large area PDs established a novel hybrid technology combining Micromegas~\cite{MM} and THick Gas Electron Multipliers (THGEMs)~\cite{THGEM_others}. 
	In 2016 COMPASS RICH-1 was upgraded by replacing four of the remaining 12 MWPCs with new detectors based on the novel hybrid MPGD  technology~\cite{upgradeHybrid}. The new detectors have been successfully commissioned and operated during the 2016 and 2017 COMPASS data taking periods.
	
	
	\section{The Architecture of the Hybrid Detector}
	
	
	\begin{figure}[h]
		\centering
		\includegraphics[width=\linewidth]{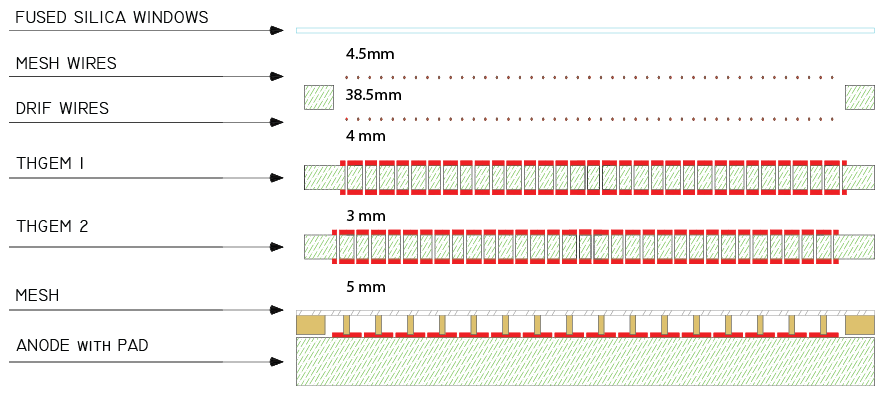}
		\caption{Schematic of the hybrid MPGD based PD (not to scale).}
		\label{fig:hybrid}
	\end{figure}
	
	The basic structure of the hybrid module (Fig.\ref{fig:hybrid}) consists of thre multiplication layers: two layers of THGEMs and a final Micromegas one. The architecture is completeded by two planes of wires. UV light sensitivity is obtained via  a thin (300 nm) CsI film deposited on the top of the first THGEM electrode which acts as a reflective photo-cathode, an approach studied also by other research groups~\cite{CsI_Breskin}. 
	\par
	The detectors are operated with $Ar:CH_{4} = 50:50$ gas mixture.
	The Drift wire plane is installed at 4~mm from the CsI coated THGEM and is biased to a suitable voltage in order to maximize the extraction and collection efficiency of the converted photo-electrons~\cite{leopard}. The other wire plane guarantees the correct closure of the drift field lines and is positioned 4.5 mm away from the quartz window which separates the radiator gas volume from the photon detector gas volume. 
	
	The photo-electron generated by the conversion of Cherenkov photon from the CsI surface is guided into one of the first THGEM holes where a first avalanche process takes place due to the electric field generated by  the biasing voltage applied between the top and bottom THGEM electrodes. The electron cloud is then driven by the electric field across the 3 mm transfer region to the second THGEM, where thanks to complete misalignment of the holes with respect to the first THGEM,
	the charge is distributed, typically, among three holes and undergoes a second multiplication process.
	
	\begin{figure}[h]
		\centering
		\includegraphics[width=0.7\linewidth]{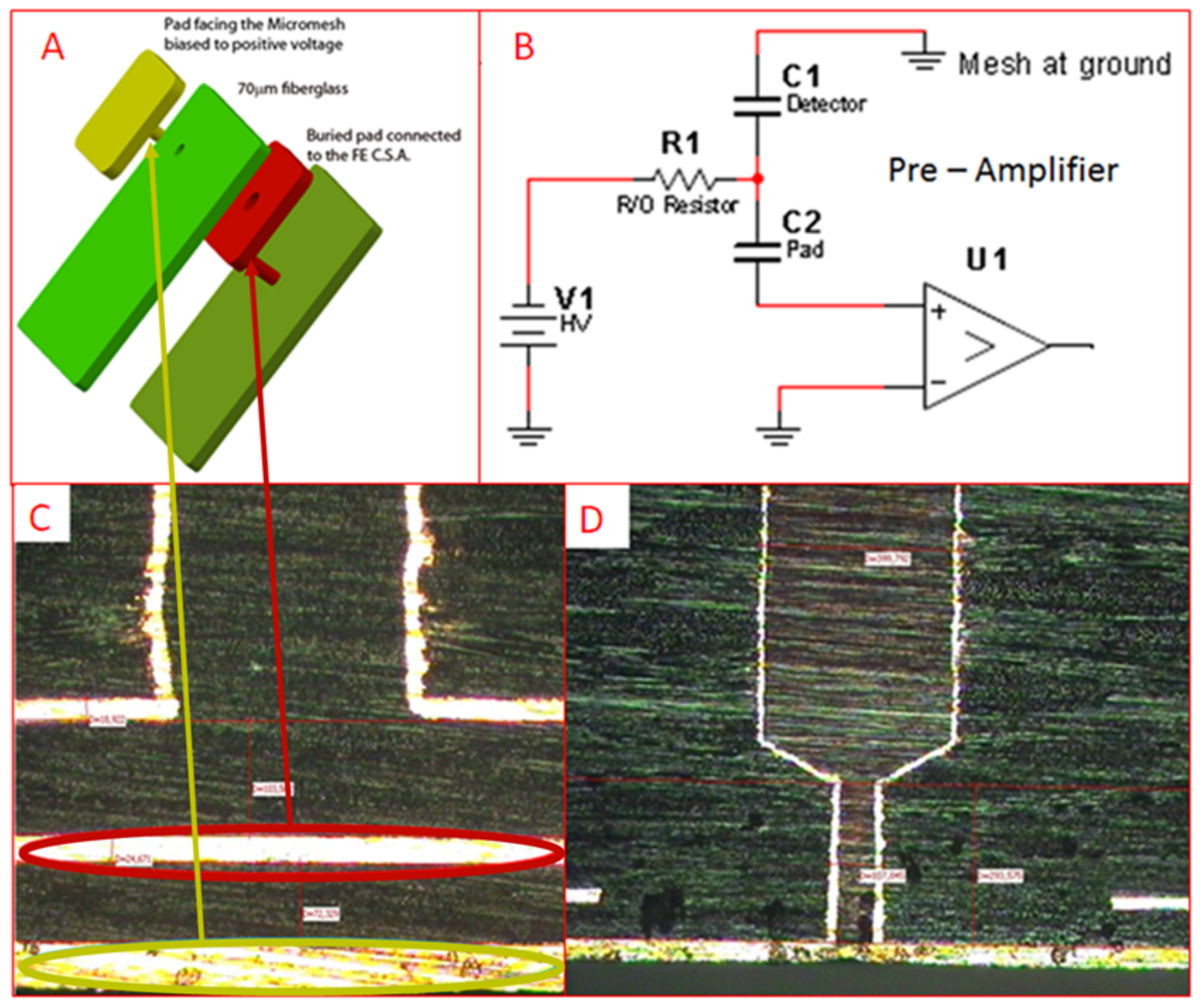}
		\caption{A: Exploded view of one single readout pad structure. B: The schematic of the circuit diagram of the resistive/capacitive anode concept. C and D: Metallographic section of the PCB: the detail of the through-via contacting the external pad through the hole of the buried pad~\cite{upgradeHybrid}}.
		\label{fig:padscheme}
	\end{figure}
	\par
	Finally the charge is guided by the electric field across the 5 mm gap to the bulk Micromegas where the last multiplication occurs. The spread of the charge over a large surface enhance the Micromegas electrical stability at higher gains. The Micromegas mesh, which is the only non-segmented electrode is kept at ground potential while the anode, segmented in square pads of 7.5$\times$7.5 $mm^{2}$ (with 0.5 mm inter-pad gaps) is biased at positive voltage (Fig.\ref{fig:padscheme}-A and -B).
	The Micromegas PCBs are based on the capacitive/resistive concept: the signal
	generated on the anode pad is capacitively transfered 
	to a second pad, parallel to the first one and burried inside the PCB. The two pads are separed by a fiberglass layer  70~$\mu$m thick. Each anode pad is powered through individual resistors. Therefore, 
	these detectors represent the first implementation in an experiment of the concept of resistrive Micromegas; this first implementation is by discrete elements, namely the resistors in series with the
	individual pads . 
	\par
	The high voltage connection to the anode pads is by vias passing through holes of the readout pads
	(Fig.\ref{fig:padscheme}-A and  -C). 
	Special attention was paid in order to obtaining a very flat surface of the anode pad via the careful
	engineering of the connections  (Fig.\ref{fig:padscheme}-D). 
	\par
	The typical voltage applied to
	the multiplication stages 
	are 1270~V across THGEM1, 1250~V 
	across THGEM2, and
	620~V to bias the MM. 
	The drift field above the first THGEM is 
	500~V/cm, the transfer field between the two THGEMs is 1000~V/cm
	and the field between the second THGEM and the MM micromesh 
	is 1000~V/cm.
	The effective gain-values
	for the three multiplication layers are 
	around 12, 10 and 120; these values include the electron transfer 
	efficiency.
	\par
	The intrinsic ion blocking capabilities of the Micromegas as well as the arrangements of the THGEM geometry and electric fields configuration
	grant an ion back flow to the photo-cathode surface lower or equal to 3\% \cite{PDreview}. Concerning the figures of ion backflow rates, this hybrid architecture has results more effective than the use of triple staggered THGEMs~\cite{our-thgem-IBF}.
	
	
	\section{Building and commissioning of the final detectors}
	
	
	\begin{figure}[h]
		\centering
		\includegraphics[width=0.7\linewidth]{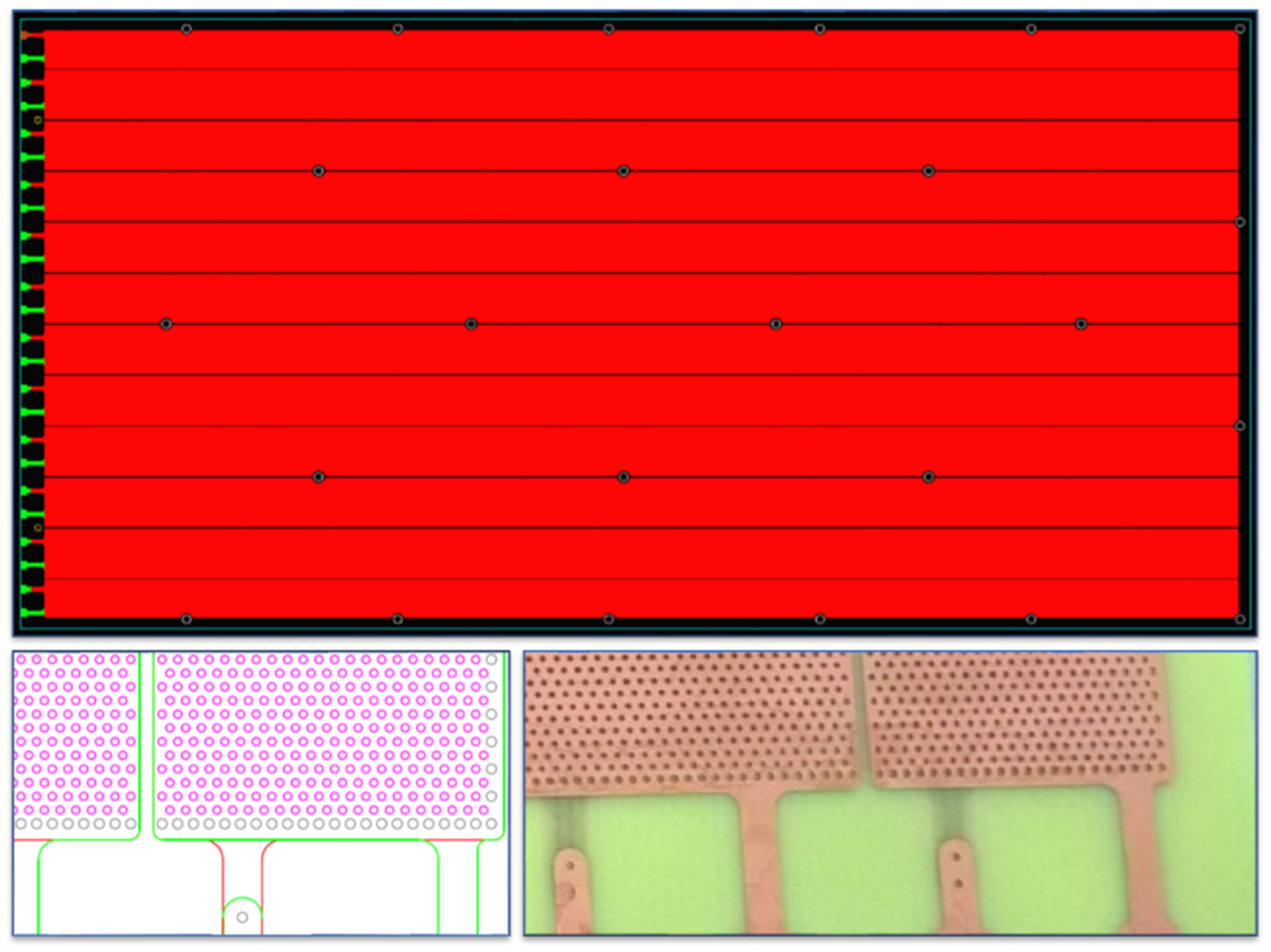}
		\caption{Top: Cad drawing of a $287\times581~mm^{2}$ THGEM with 12 sectors; Bottom left: Zoom of a corner of the CAD drawing showing the increased hole diameter of the border holes; Bottom right: Picture of the corner for a  THGEM.}
		\label{fig:thgemdrawing}
	\end{figure}
	
	\par
	All THGEMs have the same geometrical parameters: thickness of 470 $\mu$m, total length of 581 mm and width of 287 mm (Fig.\ref{fig:thgemdrawing}). The holes are arranged in a hexagonal matrix formation having diameter of 400 $\mu$m and pitch of 800 $\mu$m. The holes are produced by mechanical drilling and have no rim. To obtain a more symmetric field line configuration near the edges of the THGEM, the holes located along the external borders have 500 $\mu$m diameter. This solution provides an improved electrical stability of the whole system. The  electrodes on the two faces of each THGEM are segmented in 12 sectors separated by 0.7 mm clearance area. The biasing voltage is individually provided to each sector.
	\par
	The THGEMs are produced following a dedicated protocol, which is one of the results of an eight-year long dedicated R\&D. The protocol  includes raw material selection, THGEM production, quality assessment, characterization, CsI coating, storage and installation. 
	\par
	Achieving an effective gain uniformity  over large  areas is challenging due to the poor thickness tolerance of the raw PCB sheets available from producers. A setup based on MITUTOYO EURO CA776 coordinate measuring machine operated in a climatized room was used to map the local thickness values in order to select adequate raw material. In total 50 foils were measured from which 100 THGEMs (2 THGEMs/foils) could be produced.  Foils with a maximum thickness variation of $15$~$\mu$m 
	peak-to-peak were selected for the industrial production including transfer of the mask image, etching and drilling of the holes. 60\% of the foils have been  accepted and 60  THGEMs have been 
	industrially produced. 
	Thanks to the raw material selection the gain uniformity obtained is at the at the $\sim7\%$ level over the THGEM active surface of $\sim 0.2$~$m^{2}$.
	\par
	A post production treatment was applied at INFN Trieste to the industrially produce THGEMs: it consists in polishing the raw THGEM with pumice powder and cleaning by high pressurized water, ultrasonic bathing with high pH detergent solution ($pH \sim 11$), rinsing with demineralized water and drying in oven at 150$^{0}$C for 24 hours~\cite{Polishing}. Then, a measurement of the discharge rate was performed using an automated test setup: in an $Ar:CO_{2}$~=~$70:30$ gas mixture the bias voltage of the THGEM was increased in 10~V steps and the number of sparks (events with more than 50~nA current) was measured for 30 minutes until the bias voltage was increased again. The THGEMs with a discharge rates of less than 1/hour at 1200~$V$ were validated. Effective gain uniformity study was performed using a dedicated test setup consisting of a mini-X X-Ray generator\footnote{AMPTEK MINI-X X-Ray generator https://www.amptek.com/} and APV-25\cite{APV25} based SRS\cite{SRS}  read-out system. The best pieces were selected for the upgrade. 
	\par
	The Micromegas were produced by bulk technology~\cite{bulk-mm} at CERN EP/DT/EF/MPT workshop over the pad segmented multilayer PCBs. The $600\times600$ $mm^{2}$ PDs were built by mounting two $300\times600$ $mm^{2}$ modules side by side in the same frame.  The Micromegas PCBs have been glued to the supporting frame making use of a volumetric dispenser coupled to a computer numerical control machine. The Micromegas 
	multipliers have been assembled in a clean room.
	\par
	A special box to transport validated THGEMs under controlled atmosphere was used before and after their Au-Ni coating. The deposition of the solid photo-converter for the hybrid photo-cathodes was performed at the CERN Thin film Laboratory following the procedure described in \cite{CsI}. The photo-cathodes (THGEMs with CsI coating on one side) were mounted inside a dedicated glove-box. 
	The PDs were then installed on COMPASS RICH-1 and equipped with front-end electronics, low voltages, high voltages and cooling services.


	
	\section{The performance of the novel detectors}
	
	
	\begin{figure}
		\centering
		\includegraphics[width=\linewidth]{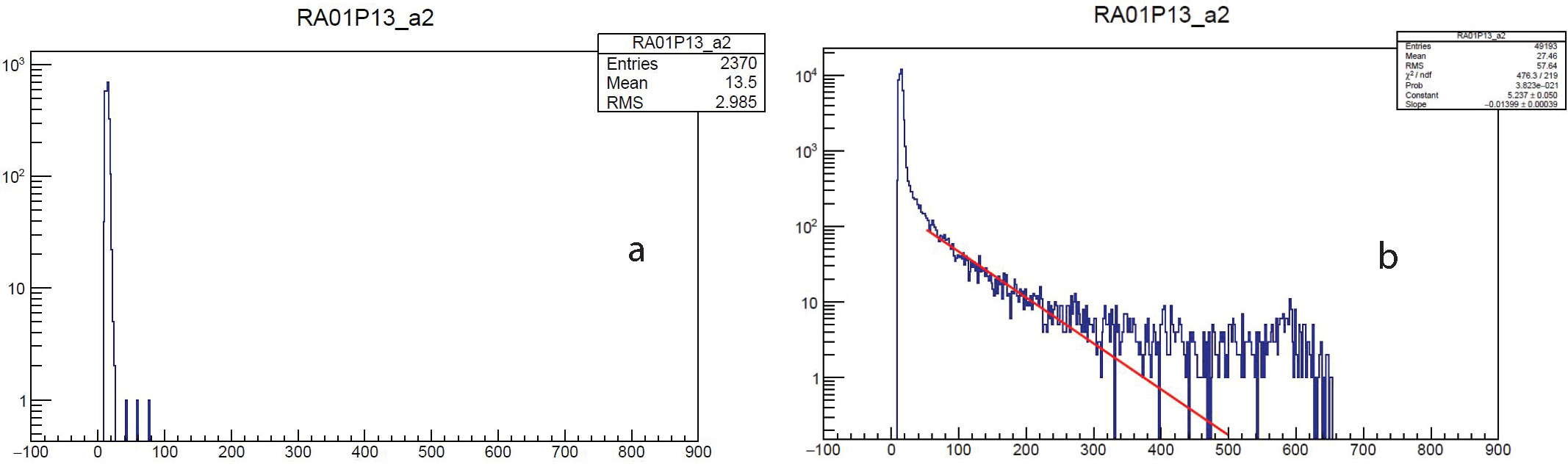}
		\caption{Amplitude spectrum from one of the hybrid PDs; each entry is the response of an anode pad with amplitude above threshold; a. No beam (random trigger); b. With beam (physics trigger).}
		\label{fig:spectrawithwithoutbeam}
	\end{figure}
	
	
	\par
	The new hybrid detectors were commissioned during the COMPASS data taking period from May to October 2016. The observed average equivalent electron noise is $\sim 900 e^{-}$. A zero suppression procedure with a $3\sigma$ threshold cut and a common mode reduction are applied for the standard data taking. After ensuring accurate timing, the amplitude spectra for noise and signals have been collected. With no beam and random trigger, the noise part of the amplitude spectrum is observed (Fig.\ref{fig:spectrawithwithoutbeam}-a). With physics triggers the amplitude spectrum shows the noise part, a prominent single photon exponential part and a tail due to charged particle signals (Fig.\ref{fig:spectrawithwithoutbeam}-b).  
	
	\begin{figure}[!htb]
		\centering
		\includegraphics[width=\linewidth]{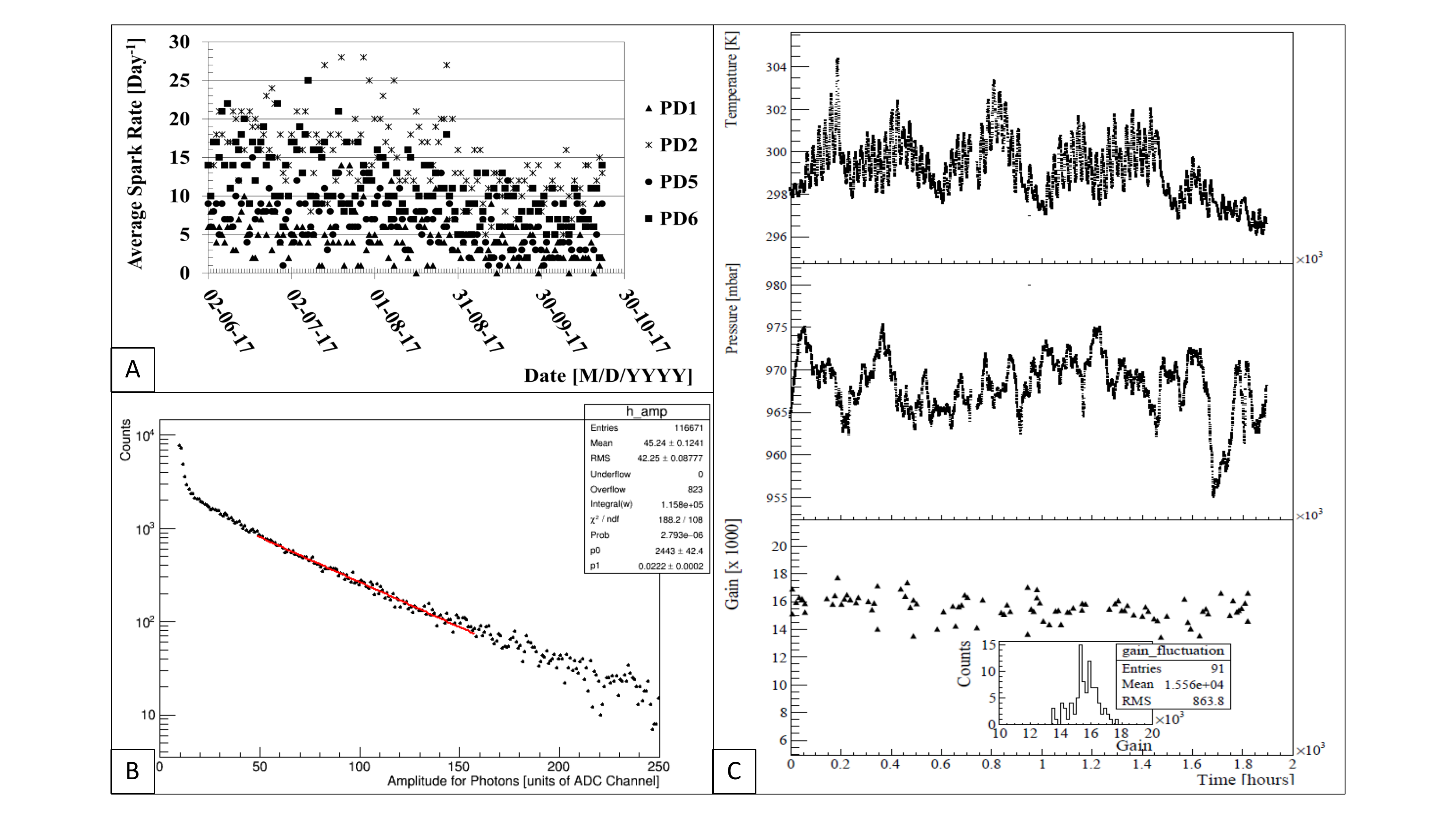}
		\caption{A. Average discharge rate per day per detector over COMPASS 2017 data taking period. B. A typical amplitude spectrum for single photons from one of the hybrid detectors with bias voltages: 1250 V for THGEM1, 1200V for THGEM2, 630 V for the Micromegas and with $1 kV.cm^{-1}$ transfer fields between amplification stages. The estimated effective gain is $13511.4 \pm 121.7$ for a fit range of 50 to 160 ADC channels. C. Temperature, Pressure and effective Gain fluctuations over 2017 COMPASS data taking period; two data points per day collected at 6.00`am and 6.00~am, namely when the daily temperature is expected minimum and maximum respectively.} 
		\label{fig:performance}
	\end{figure}
	
	\par
	Important gain variation with environmental conditions
	(pressure and temperature) are expected in a multilayer
	gaseous detector. During the R\&D phase, these effects had been studied in order to introduce  an automated high voltage correction system to compensate for the effect and to stabilize the PD gain response~\cite{hybrid-hv}. During the data taking,
	the performance, in terms of voltage and gain stability of the PDs have been studied. In Fig.\ref{fig:performance}-A. the average spark rate per day for each of the hybrid detectors is provided for the COMPASS 2017 data taking period. A typical discharge rate of few events per day is seen. No sizable voltage drop is observed when a discharge occurs, confirming the validity of the design of the HV distribution system~\cite{hybrid-hv}. In Fig.\ref{fig:performance}-B. a typical signal amplitude spectrum has been shown. The estimated gain is around 13.5k. 
	In Fig.\ref{fig:performance}-C. the measured PD gas pressure and temperature curves are shown. The effective gain stability can be extracted from Fig.\ref{fig:performance}-D, where the values of the effective gains extracted from the data, two measurements per day, at the time when the minimum and maximum temperature is expected, are presented. The gain is stable at the 6\% level. 
	

	
	\begin{figure}[!hbt]
		\centering
		\includegraphics[width=\linewidth]{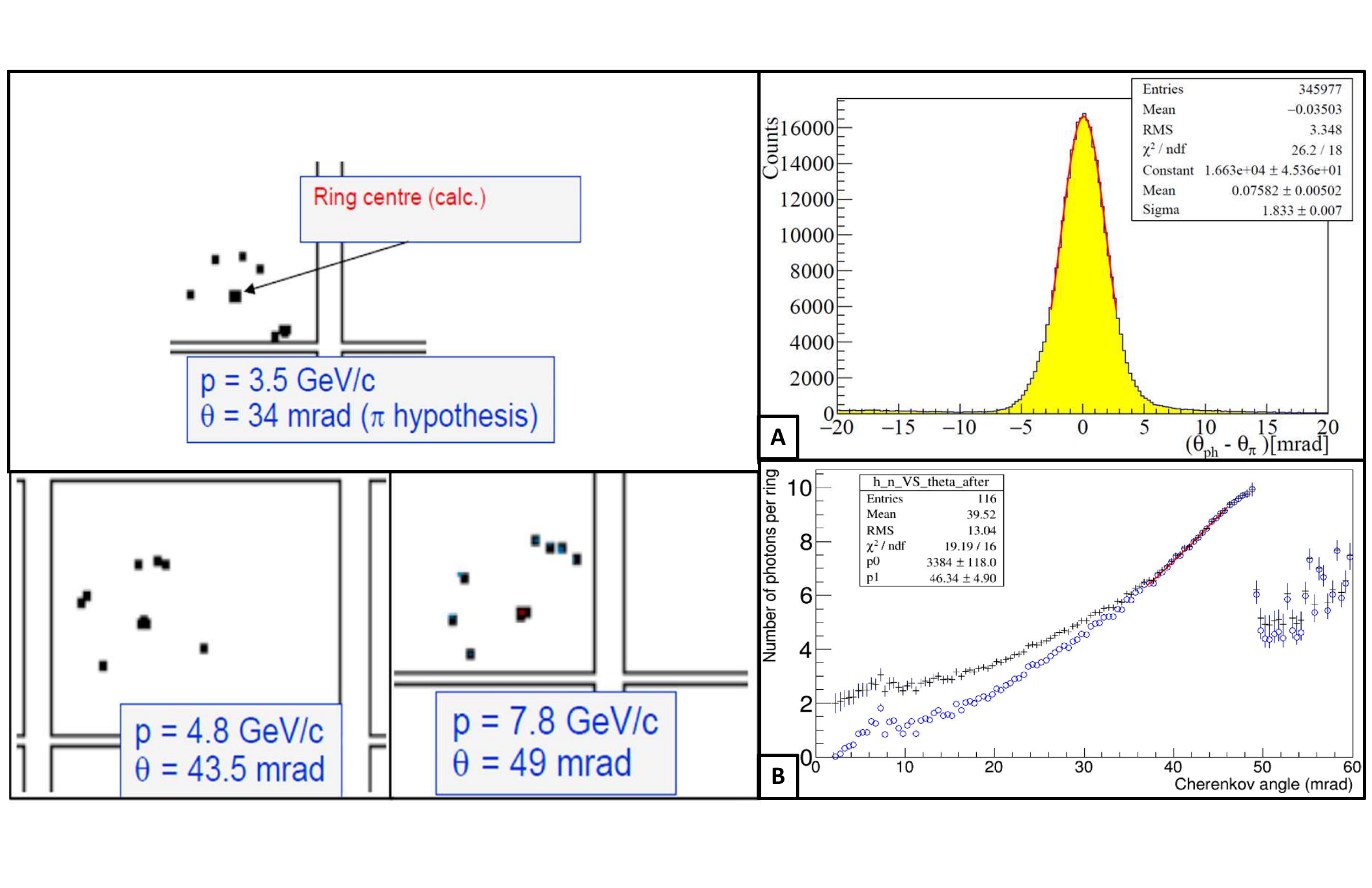}
		\caption{Left: Images of hit pattern in the novel photon detectors. The center of the expected ring patterns is obtained from the reconstructed particle trajectories; the particle momentum and the expected Cherenkov angle in the pion hypothesis are also reported. No image elaboration or background subtraction is applied. Right, A: Distribution of the difference between the Cherenkov angle calculated from the reconstructed particle momentum and the Cherenkov angle provided by single detected photoelectrons; a sample of identified pions is used. Right, B: Average number of detected single photons per Cherenkov rings versus the Cherenkov angle.}
		\label{fig:ringsPh}
	\end{figure}
	
	\par
	Typical Cherenkov rings are presented in Fig.\ref{fig:ringsPh}-left. The centres of the expected ring patterns is obtained from the reconstructed particle trajectories; the particle momentum and the expected Cherenkov angle in the pion hypothesis are also reported. No image elaboration or background subtraction is applied.
	The analysis shows that the residual angular resolution of those PDs are $\sim$~1.86 mrad (Fig\ref{fig:ringsPh}-A.). To estimate the number of photons per Cherenkov rings the fit function Eqn\ref{eq:fitCherenkovFunc} has been used: 
	
	\begin{equation}
	\centering
	N_{\theta_{ch}}=p_{0}.sin^{2}\theta_{ch}+p_{1}.\theta_{ch}
	\label{eq:fitCherenkovFunc}
	\end{equation}
	
	where $N_{\theta_{ch}}$ is the average number of detected photons per Cherenkov ring for a particular Cherenkov angle $\theta_{ch}$, $p_{0}$ and $p_{1}$ are the fit parameters for the Cherenkov photon part according to the Frank and Tumm distribution and for the background part of the data respectively, assumed linear as suggested by geometrical considerations. The fit estimation indicates 12.9 photons per ring  at Cherenkov angle equals to 55.2 mrad; the signal part is $10.3$ $\pm$ $0.4$ and background part is $2.6 \pm 0.3$ (Fig.\ref{fig:ringsPh}-B).
	

	\section{Future prospectives}
	
	\par
	The THGEM-MM hybrid PD technology could be used for other applications, for instance for a RICH dedicated to PID of high momentum particles at experiments at the future EIC collider \cite{EIC}, where higher space resolution of the PD would be needed because of the more compressed geometry that imposes a reduced lever arm.
	\par
	A prototype \cite{NanoDiamond} similar to the COMPASS PDs has been designed and built: it has an active area of 10$\times$10 cm$^{2}$,	an anode segmented in 1024 square pads having 3.5 mm pitch. The prototype is fully	modular, with front-end electronics and almost all services contained in the 10$\times$10 cm$^{2}$ active area:	detectors covering larger areas could be designed by multiple replica of the basic module represented by the prototype.
	
	
	\begin{figure}
		\centering
		\includegraphics[width=0.9\linewidth]{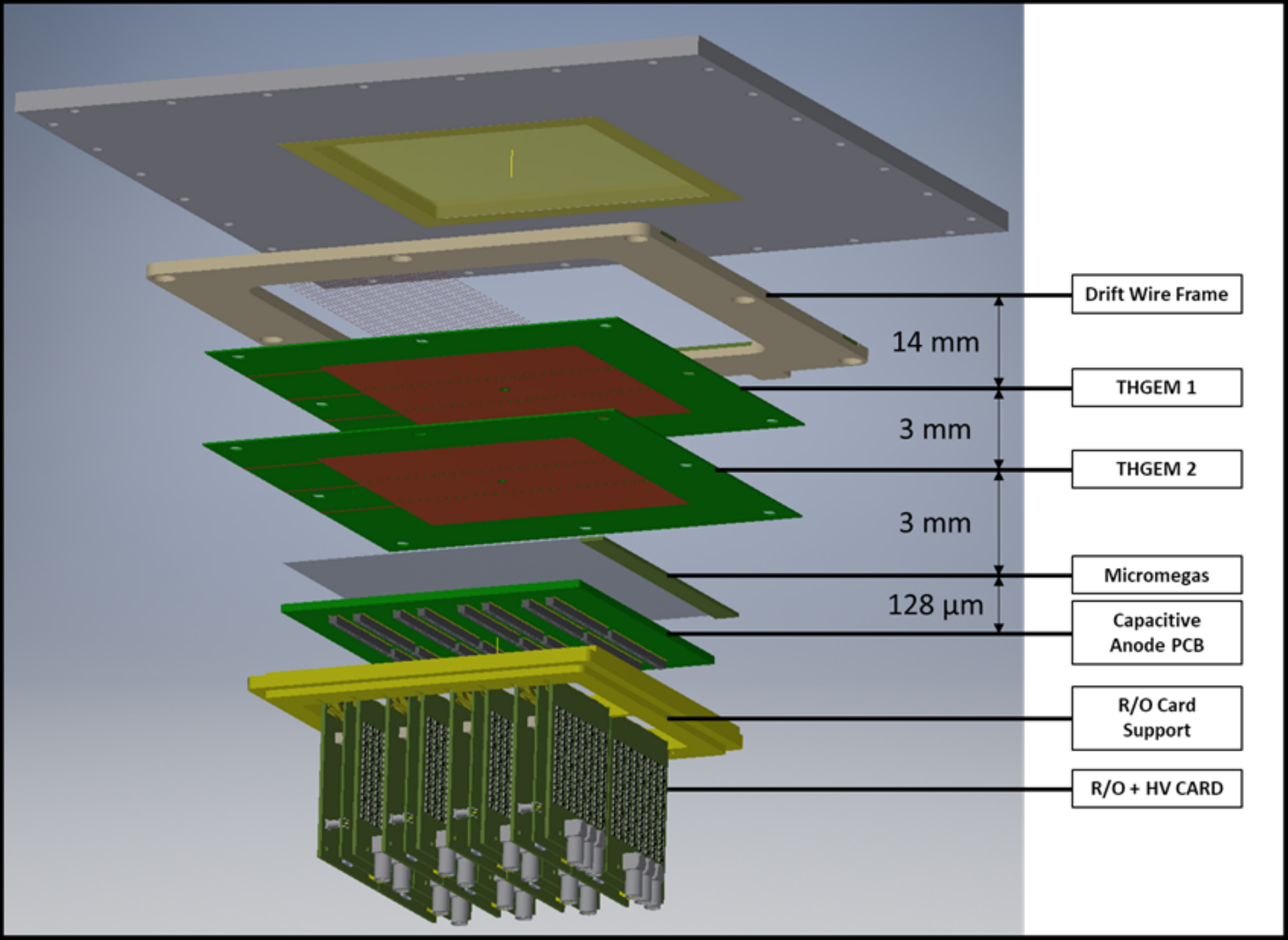}
		\caption{Exploded view of Mini-PAD hybrid schematic from CAD drawing}
		\label{fig:hybridminipadschematic}
	\end{figure}
	
	The internal structure of the modular minipad prototype presented in Fig.\ref{fig:hybridminipadschematic} 
	and it reproduces the basic scheme of the hybrid MPGD implemented for COMPASS RICH-1.
	
	The detector components have been prepared adopting procedures and protocols similar to those used for COMPASS hybrid PDs. The prototype has been tested at INFN Trieste laboratories  and, in October - November 2018,  in a test beam at CERN SPS.  The beam particles, alternatively $\pi$ and $\mu$, traversed on a truncated cone fused silica Cherenkov radiator aligned with the center of the pad plane. A shutter, remotely controlled, sitting between 
	the radiator and the detector, made possible to collected data including Cherenkov photons from the radiator or
	excluding them.
	
	
	\begin{figure}
		\centering
		\includegraphics[width=1.0\linewidth]{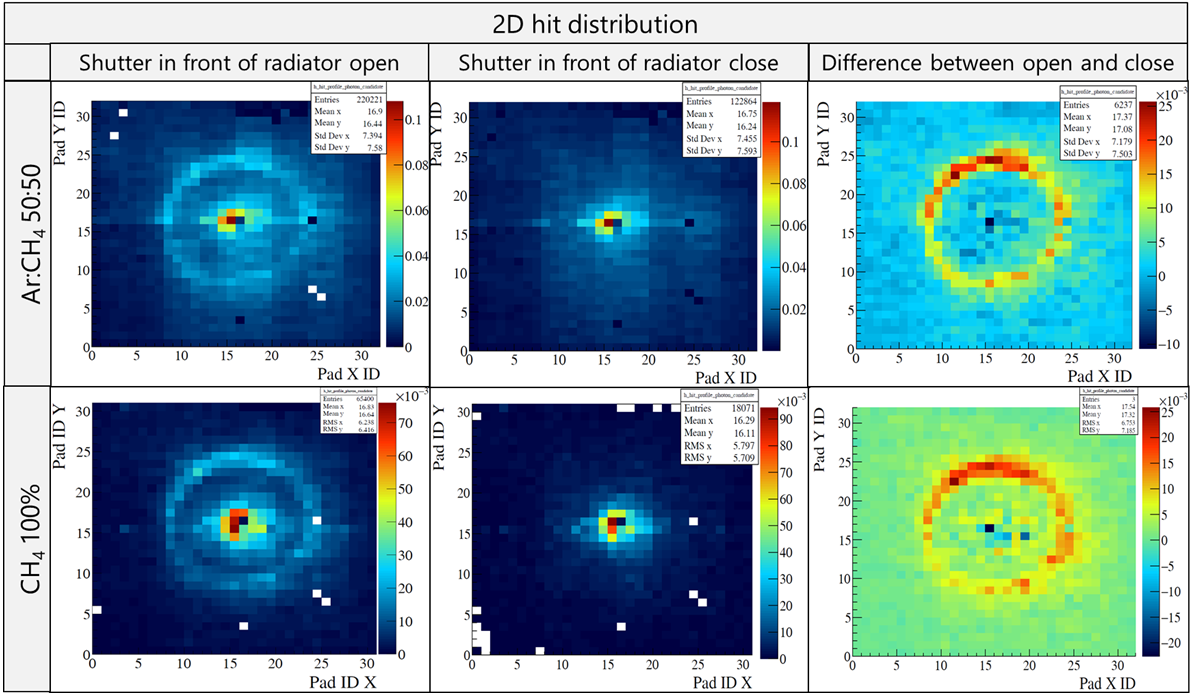}
		\caption{Observed Cherenkov rings in $Ar:CH_{4}$ 50:50 gas mixture and in pure $CH_{4}$ gas. First column: data collected with the shutter open. Central column: data collected with the shutter closed. Right column:
			distributions obtained subtracting those obtained with shutter open and shutter closed after properly nomalization to the same number of triggers.}
		\label{fig:CerenkovRings}
	\end{figure}
	
	Cherenkov rings has been observed for different $CH_{4}$ rich gas mixtures as shown in Fig.\ref{fig:CerenkovRings}.
	The analysis of the test beam data is ongoing.
	

	\section{Conclusions}
	
	\par
	Novel MPGD-based hybrid detectors of single photons, matching the present status of the art in the MPGD technology,  have been designed, engineered, built, tested and operated in a RICH detector for the first time. The main achievements are:
	\begin{itemize}
		\item The four new PDs have been stably operated during 2017 COMPASS data taking period matching the COMPASS RICH requirements. 
		\item It is the first time that single photon detection is performed in an experiment by MPGD PDs.
		\item It is the first application in an experiment making use of THGEMS and resistive Micromegas.
		\item The typical gain, well above 10$^4$, is the highest at which MPGD have been operated in experiments. 
		\item The characterization shows the expected angular resolution and around 10-12 detected photons per ring at Cherenkov angle saturation.
		\item The hybrid Micromegas and THGEM photon-detection technology has proven to be successful and robust.
		\item There are promising perspective for the increase of the spacial resolution using this technology.
	\end{itemize}
	A new R\&D has already been started for the future needs of the EIC experiments with modified design parameters to achieve a higher space resolution.
	
	
	

	\section{Acknowledgment}
	\par
	The activity is partially supported by the H2020 project AIDA2020 GA no. 654168. 
	\par
	The authors are member of the COMPASS Collaboration and part of them are members of the RD51 Collaboration: they are grateful to both Collaborations for the effective support and the precious encouragements. 
	\par
	The Portuguese collaboration was partially supported by CERN\slash~FIS-PAR\slash0022\slash2019 through~FCT~(Lisbon). 
	

	\section{References} 
	
\end{document}